\documentclass[aps,preprint]{revtex4}
\usepackage{graphicx}
\usepackage{rotating}
\usepackage{dcolumn}
\usepackage{amssymb}
\usepackage{amsmath}
\usepackage{epsfig}

\begin{document}

\title{Conformal Universality in Normal Matrix Ensembles}
\author{A.M. Veneziani$^1$, T. Pereira$^{2}$, and D.H.U. Marchetti$^1$}
\affiliation{ $^1$Institute for Physics, University of S\~ao Paulo, S\~ao Paulo, Brazil \\
 $^2$Center For Mathematics, Federal University of ABC, S\~ao Paulo, Brazil}
\date{\today}

\begin{abstract}
	A remarkable property of Hermitian ensembles is their universal behavior,  
	that is, once properly rescaled the eigenvalue statistics does not depend on 
	particularities of the ensemble. Recently, normal matrix ensembles have 
	attracted increasing attention, however, questions on universality for these 
	ensembles still remain under debate. We analyze the universality properties 
	of random normal ensembles. We show that the  concept of universality  
	used for Hermitian  ensembles cannot be directly extrapolated to normal 
	ensembles. Moreover, we show that the eigenvalue statistics of random
	 normal matrices with radially symmetric potential can be made universal 
	 under a conformal transformation. 
\end{abstract}

\maketitle

One of the most celebrated results in random matrix theory is the universal
behavior of the eigenvalues statistics of Hermitian ensembles. The
universality states that the eigenvalues statistics considered on the proper
scale (with unit the mean eigenvalue spacing) is always the same,
independently on the particularities of the ensemble. The universality in
random matrix theories plays a major role in many branches of 
sciences \cite{Haake,Nuclear,Mehta,QCD}.

The universality problem is addressed by using the so-called 
\textit{invariant ensemble model}. The probability  
\begin{equation}
P(M)dM\propto \exp \left\{ -n\text{Tr}[V(M)]\right\} dM,  \label{Pm}
\end{equation}
of finding a $n\times n$ matrix $M$ within the ensemble, with $dM$ the
Riemann volume in the space of normal matrices, is invariant by unitary
transformations. The corresponding eigenvalue density, in the limit 
$n\rightarrow \infty $, depends on the particular form of $V(M)$.

The universality in the eigenvalue statistics can be heuristically understood in 
terms of the Dyson interpretation of the eigenvalues as a Coulomb gas 
\cite{Mehta}. For a hermitian matrix Eq. (\ref{Pm}) can be written in terms of the 
joint probability of the eigenvalues 
{\small
\begin{equation}
P_{n}(\lambda_{1},\cdots ,\lambda_{n})\propto \exp 
\Big\{ \sum_{ 1 \le i\neq j\leq n }\ln | \lambda_{i}-
\lambda_{j} |^{-1} - n\sum_{i=1}^{n}V(\lambda_{i}) \Big\}.  
\label{Pe3}
\end{equation}}
The joint distribution of eigenvalues has the same form as a 2 dimensional 
Coulomb gas restricted to one dimension.  The eigenvalue interaction is given 
by a Coulomb term $ \ln | \lambda_i -  \lambda_j |^{-1}$ under a 
potential $V(\lambda)$ that confines the eigenvalues in a bounded set of the 
real line.

Since the eigenvalues are restricted to one dimension,  if an eigenvalue is 
located between two eigenvalues the Coulomb potential  does not 
allow it to switch its position with the surrounding eigenvalues. As the number 
of eigenvalues increase (the size of the matrix) the eigenvalues become closer 
and the effect of the Coulomb potential overtakes the potential $V(\lambda)$ 
(due to the logarithm behavior near zero).

If we then rescale the mean distance between eigenvalues to the unity,
heuristically this is equivalence to \textit{switch off} the external
potential. Thus no matter the external potential, after the rescaling only
the Coulomb potential affects the eigenvalues. Recently, all this reasoning
has been rigorously proven (see e.g.  \cite{Deift} and references therein).

Over the last decade, normal matrix ensembles have attracted a great 
deal of attention, mainly due to their broad applications such as in quantum 
Hall effect \cite{Hall}, quantum flows \cite{Flows}, quantum field theory \cite{Feinberg}, 
conformal hierarchies and  conformal maps \cite{Zabrodin,Wiegman}.  
Despite of the increasing interest on normal ensembles, questions on their 
universality properties still remain open.

The joint probability of the eigenvalues of a normal ensemble obeys an 
equation  of the type of Eq. (\ref{Pe3}). The basic difference is the dimension 
of the eigenvalues $z_i$.  For normal matrices the eigenvalues lie on the complex plane. 
The different eigenvalue dimensions between Hermitian and normal ensembles 
may lead to different universal behaviors. In the later case, as we add 
eigenvalues they can turn around and occupy a variety of positions in the 
complex plane to  finally be set to  minimize the electrostatic energy. It is 
possible that after this process the rescaling of the average  distance still 
brings out the particularities of the potential. If so, what would be the procedure 
to assure the universal behavior in normal ensemble, if any? 

In this letter, we show that ensembles of normal matrices under radially
symmetric potential are not universal in the Hermitian sense, that is, if
one rescales the mean distance to unity the eigenvalue statistics still
depends on the potential. We then show that the universal
properties can be retrieved by conformal transformations of the rescaled
eigenvalue distribution.

Recently, questions on universality of normal ensembles have been addressed 
in Ref. \cite{Chau}. The machinery of statistical mechanics was used to argue 
that normal ensembles presents a {\it universality of the second type} \cite{Zee}. 
This means that  the one-point Green's function and the connected two-point 
Green's functions may be related to each other in a universal way after some 
proper normalization and rescaling of the system.  Yet, such proper rescaling 
remains unknown, and even more serious question remains: whether universality 
of the second  type is the only kind of universality ruling normal ensembles 
\cite{comment}. 
 
\textit{Hermitian ensembles:} Let us first revise the precise meaning of
universality for Hermitian ensembles. The quantity playing a major role in
the analysis of random ensembles is the integral kernel 
\begin{equation}
K_{n}\left( x,y\right) =e^{-\frac{n}{2}\left( V(x)+V(y)\right)
}\sum_{j=1}^{n}\phi _{j}\left( x\right) \phi _{j}\left( y\right) 
\label{KuN}
\end{equation}
where $V(x)$ is the potential [see Eq. (\ref{Pe3})] and $\left\{ \phi
_{j}\right\} _{j=1}^{n}$ denotes the set of polynomials up to order $n-1$,
orthogonal with respect to the weight $\exp \{-nV(x)\}$. The important 
statistical quantities associated with the matrix ensemble such as the 
eigenvalue density $\rho (\lambda _{1})=\int
P_{n}(\lambda _{1},\cdots ,\lambda _{n})d\lambda _{2}\cdots \lambda _{n}$
can be obtained by the $K_{n}$ \cite{Deift}. For large $n$ the relation reads 
$$
K_{n}\left( \lambda ,\lambda \right) =n\rho (\lambda )\left( 1+o\left(
1\right) \right) ,
$$ 
where $o(1)$ converges to $0$ as $n\rightarrow \infty $.

To proceed the universality analysis,  we must rescale the integral kernel. The 
scale is chosen such that the fraction of eigenvalues in an interval of length $s$ 
close to a point $\lambda$ equals $s$, in other words, the average spacing 
between eigenvalues is unity. It can be shown that the correct scale is 
$K_n (\lambda, \lambda)$. Indeed, since the fraction of eigenvalues in the interval 
$A \subset \mathbb{ R}$ is given by 
\begin{equation}
f_{n} ( {A} ) =\frac{1}{n}\sum_{i=1}^{n}
\chi _{{A}} (\lambda _{i} ),
\end{equation}
it is possible to show that, for large $n$, it holds 
$ \langle n f_{n} [ \lambda ,\lambda + s/K_{n}( \lambda ,\lambda ) ] \rangle =
s[ 1+o( 1) ]$ \cite{DKMVZ}.  Once we have the proper scale 
$K_{n}( \lambda ,\lambda )$, we proceed the analysis by rescaling the integral 
kernel 
{\small
\begin{equation}
\widetilde{K}_{n} ( x,y) =\frac{1}{K_{n}( \lambda, \lambda) }K_{n}
\left( \lambda +\frac{x}{K_{n} ( \lambda,\lambda) },\lambda +\frac{y}
{K_{n}( \lambda ,\lambda) }\right).
\label{rk}
\end{equation}}

The astonishing result \cite{DKMVZ} is then 
\begin{equation}
{\lim_{n\rightarrow \infty }}~\widetilde{K}_{n}\left( x,y\right) =\frac{\sin
\pi (x-y)}{\pi (x-y)}
\end{equation}
exists pointwise for Freud--type or real analytic potentials $V$. Since it
does not depend on $V$ the Hermitian ensembles in this sense are universal.
The natural questions is whether normal ensembles also display universal
eigenvalue statistics.

{\it Radially symmetric normal ensembles:} We shall adapt these questions on 
universality to the family of normal ensembles. We shall focus on the case of 
normal ensembles with a radial potential
\begin{equation}
V_{\alpha }\left( z\right) = | z | ^{\alpha }.
\label{Valpha2}
\end{equation}
we shall consider the potential for $\alpha \ge 2$. The statistical quantities of this
ensemble can be characterized by using the theory of logarithmic potentials and 
orthogonal polynomials developed in Ref. \cite{Totik}. The eigenvalue density 
associated with the potential $V_\alpha(z)$ is given by $ \rho (z) = \alpha^2 
|z|^ {\alpha - 2}/ 4 \pi $, the support $supp(\rho)$ of the eigenvalue density can 
be computed by the condition $|z| V^{\prime}_{\alpha}(|z|) \le 2$, which  yields 
$$
supp(\rho) = \{ z \in \mathbb{C} : |z|  \le (2 / \alpha)^{1/ \alpha} \}.
$$ 

The integral kernel associated with $V_{\alpha}$ is then given by 
$K_{n}^{\alpha} ( z,w ) = \exp \{- \frac{n}{2} [ V_{\alpha} (z) + 
\overline{V_{\alpha}(w)} ] \} \sum_{j=1}^{n} \phi_{j}(z) \overline{\phi_{j}(w)} $, 
where $\phi _{j} ( w )$ are orthogonal polynomials with respect to the weights  
$\exp\{ n V_{\alpha}(z)\}$. Since, $V_{\alpha}(z)$ is radially symmetric the orthogonal 
polynomials $\phi _{j} ( w )$ are monomials, using the explicit relations for 
$\phi_j(z)$ we obtain
\begin{equation}
\frac{1}{n}K_{n}^{\alpha } ( z,w ) = \frac{\alpha e^{-\frac{n}{2} | z |^{\alpha }}e^{-\frac{n
}{2} | w | ^{\alpha }}}{2\pi  z\overline{w} } \sum_{j=1}^{n}\frac{n^{\frac{2j}{\alpha }-1} 
( z\overline{w} ) ^{j}}{\Gamma 
\left( \frac{2j}{\alpha }\right)}
\label{kna}     
\end{equation}   

\begin{figure}[!h]
\centerline{\hbox{\psfig{file=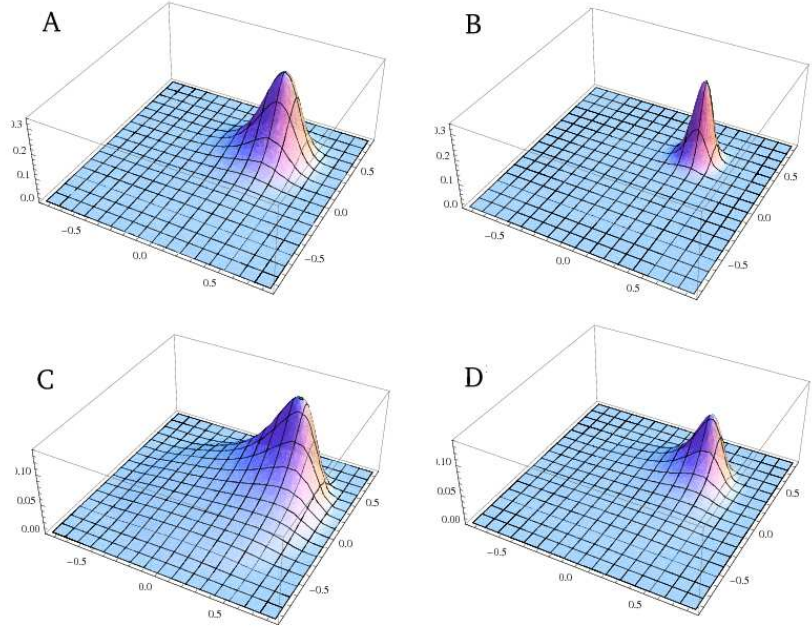,height=9cm}}}
\caption{ Profile of the integral kernel $n^{-1} \Big| K_n^{\alpha}(z,w) \Big|$ for 
fixed $z=0.3 + i 0.4$ as a function of 
$w=x+iy$. We show distinct values of $\alpha$ and $n$. One can see that 
the kernel presents a peak that is sharped with the increase of $n$. This 
behavior persists for all $\alpha \ge 2$. The distinct values of $\alpha$ are 
$\alpha=4$ in (A,B) and  $n=50, 200$ and $\alpha=8$ for $n=50, 200$ in (C,D).}   
\label{Fig1}
\end{figure}

For $n$ large enough $K_{n}^{\alpha }( z,w )$ can be made small everywhere, 
except  in a vicinity where $K_{n}^{\alpha } ( z,w )$ attains its maximum, which 
occurs when $z=w$. In Fig. (\ref{Fig1}), for $z = 0.3 + i 0.4$ we depict 
$n^{-1} \Big| K_n^{\alpha}(z,w)\Big|$ as a function of 
$w= x + iy$. One can clearly see that as $n$ increases the peak is sharped 
at $w=z$.  $n^{-1} \Big| K_n^{\alpha}(z,w) \Big|$ presents the same behavior 
for different values of $z \in supp(\rho)$.

\textit{Asymptotic expansion for the integral kernel:} We wish to obtain an
asymptotic expression for the integral kernel $K_{n}^{\alpha }$. We are
concerned with the behavior of the integral kernel for all $w$ in a vicinity
of $z$, only the values of $w$ close $z$ gives a relevant contribution. We
shall investigate the behavior of $K_{n}^{\alpha }$ in the region 
\begin{equation*}
\Delta _{\alpha }:=\Big\{ z,w \in supp(\rho) : \left\vert
\arg z-\arg w\right\vert <\frac{2\pi }{\alpha } \Big\},
\end{equation*}
where $arg z$ stands for the argument of the complex number $z$.
We shall explain this particular choice latter on. For now it is important
that $\Delta _{\alpha }$ contains the peak of $K_{n}^{\alpha }(z,w)$.

In the vicinity $\Delta _{\alpha }$ we can approximate Eq. (\ref{kna}) by an
integral and then solve it by means of the steepest descent method.
Proceeding in this way, we obtain an asymptotic expansion for the integral
kernel. The technical details of this analysis shall be given elsewhere.  
The asymptotic expansion reads 
{\small 
\begin{equation}
n^{-1}K_{n}^{\alpha }(z,w)=\left\{ 
\begin{array}{lll}
n^{-1} \widehat{K}_{n}^{\alpha }(z,w)\Big(1+o(1)\Big) &  & \text{if }(z,w)\in
\Delta _{\alpha } \\ 
0 &  & \text{otherwise}
\end{array}
\right. 
\label{as}
\end{equation}
}with the leading part of the asymptotic expansion being{\small 
\begin{equation*}
\frac{1}{n}\widehat{K}_{n}^{\alpha }(z,w)=\frac{\alpha ^{2}}{4\pi }(z
\overline{w})^{\frac{\alpha }{2}-1}\exp \Big\{n[(z\overline{w})^{\frac{
\alpha }{2}}-\frac{|z|^{\alpha }}{2}-\frac{|w|^{\alpha }}{2}]\Big\}.
\end{equation*}
} \noindent
Now it becomes clear the behavior seen in Fig. (\ref{Fig1}). Since $2\mbox{Re}(z
\bar{w})^{\alpha /2}\leq |z|^{\alpha }+|w|^{\alpha }$, being equal only if $
z=w$,  with the increase of $n$, $K_{n}^{\alpha }$ converges to zero
exponentially fast whenever $z\not=w$.
 
\begin{figure}[!h]
\centerline{\hbox{\psfig{file=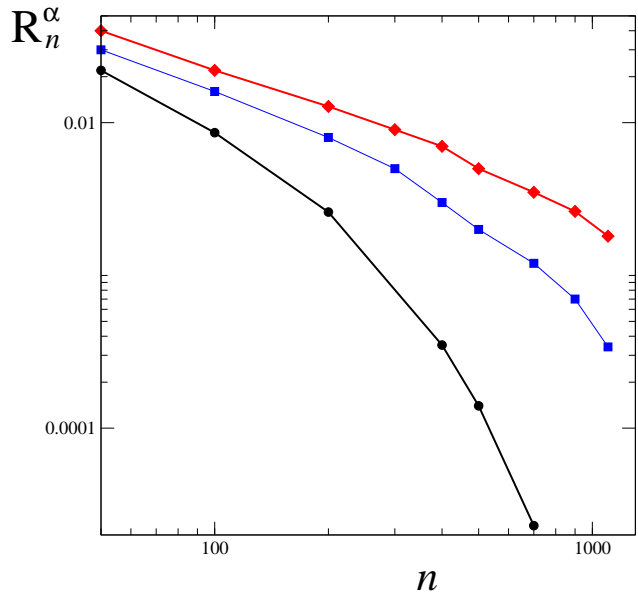,height=9cm}}}
\caption{ We depict the maximum difference $R_n^{\protect\alpha}$ versus 
$n$. We plot $R_n^{\protect\alpha}$ for three
distinct values of $\alpha$, namely $\protect\alpha=6$ (circles), 
$ \alpha=8$ (squares) and $\protect\alpha = 11$ (diamonds). One can
clearly see the $o(1)$ behavior of $R_n^{\alpha}$. }
\label{FigError}
\end{figure}

The $o(1)$ behavior of the asymptotic expansion can be seen by
introducing the quantity 
\begin{equation*}
R_n^{\alpha} = \displaystyle \max_{z,w \in supp(\rho) } \left| n^{-1}K_n^{\alpha}(z,w) - 
n^{-1} \widehat{K}_n^{\alpha}(z,w) \right|, 
\end{equation*}
which gives the largest value of the difference between the two
kernels in $\Delta_{\alpha}.$ In Fig. (\ref{FigError})  we depict 
$R_n^{\alpha}$ as a function of $n$ for
different values of $\alpha$. One can clearly see the
behavior $o(1)$ of $R_n^{\alpha}$. 

Outside the region $\Delta_{\alpha}$ the quantity 
$n^{-1} \widehat{K}_{n}^{\alpha }$ may differ from $n^{-1} K_n^{\alpha}$. 
$ | \widehat{K}_{n}^{\alpha }(z,w) | $ has a term $\exp \{n(z\bar{w})^{\alpha /2}\}$, 
whose absolute value is a periodic function of period of the argument 
$(z\bar{w})$. To see this we write again $z=re^{i\theta }$ and 
$w=se^{i\xi }$, which yields 
\begin{equation*}
\left\vert \exp \{n(z\bar{w})^{\alpha /2}\}\right\vert =\exp \{n(rs)^{\alpha
/2}\cos [\alpha (\theta -\xi )/2]\}.
\end{equation*}
As a consequence other peaks may occur in an angular distance $4\pi /\alpha $ 
far apart from the main peak localized in $\Delta _{\alpha }$.

\textit{Rescaled integral kernel:} As for the Hermitian ensembles, we shall
rescale the average spacing between eigenvalues. Denoting by $D_{s}$ the
disk of radius $s$ in the neighborhood of the origin, we must find a scale 
$g(s)$ such that the fraction of the eigenvalues in $D_{s}$ equals the area
of the disk. We must have $\langle nf_{n}({D}_{g(s)})\rangle =\pi
s^{2}[1+o(1)]$, with $f_{n}$ the indicator function as defined before.
Explicitly, for large $n$ through the equilibrium density $\rho (re^{i\theta
})$, using $z=re^{i\theta }$ we have 
\begin{equation*}
\Big\langle nf_{n}({D}_{g(s)})\Big\rangle=n\int_{0}^{2\pi
}\int_{0}^{g(s)}\rho (re^{i\theta })rdrd\theta \left( 1+o\left( 1\right)
\right) .
\end{equation*}
solving the integral it yields $\frac{n\alpha }{2}\left( \left( g(s)\right) ^{\alpha }\right) \left(
1+o(1)\right) =\pi s^{2}[1+o(1)]$. The scaling function has the form 
\begin{equation}
g\left( s\right) =\left( \frac{2\pi s^{2}}{n\alpha }\right) ^{\frac{1}{\alpha }}.  
\label{gas}
\end{equation}

Considering only terms of superior order in $n$ the rescaled integral kernel
is given by 
\begin{equation}
\widetilde{K}_{\alpha }\left( z,w\right) = g ^{\prime }\left( z\right) 
\widehat{K}_{n}^{\alpha }\left( g \left( z\right) ,g \left( w\right) \right) 
\overline{g ^{\prime }\left( w\right) }.  \label{rka}
\end{equation}

Note that the difference between $\widetilde{K}_{\alpha }(z,w)$ and the
rescaled kernel for the Hermitian ensemble in Eq. (\ref{rk}) is due to the
Jacobian of the coordinate change of $w\mapsto g(w)$ in the complex plane.
After some manipulations the rescaled kernel reads 
\begin{equation}
\widetilde{K}_{\alpha }(z,w)=\frac{2}{\alpha }e^{\frac{2\pi }{\alpha }\left(
z\overline{w}-\frac{|z|^{2}}{2}-\frac{|w|^{2}}{2}\right) }.
\end{equation}

Hence, the rescaled integral kernel, unlike the rescaled kernel for
Hermitian ensembles, still carries peculiarities of the potential.
Therefore, the concept of universality used for Hermitian ensembles cannot
be directly extrapolated to normal ensembles. The question then turns to
whether there is a scheme to obtain a universal integral kernel and
consequently a universal eigenvalue statistics.

\textit{Conformal Universality:} In the following, we shall show that there
is a conformal map $\varphi $ such that 
\begin{equation*}
\widehat{K}_{n}^{\alpha }\overset{\varphi }{\longrightarrow }K,
\end{equation*}
where $K$ does not depend on the potential $V_{\alpha }$ and, in this sense,
is a universal kernel. The universal kernel 
\begin{equation}
K(z,w)=\frac{1}{\pi }e^{z\overline{w}-\frac{\left\vert z\right\vert ^{2}}{2}-%
\frac{\left\vert w\right\vert ^{2}}{2}}.
  \label{Kun}
\end{equation}
The conformal map $\varphi :\Sigma _{\alpha }(z)\longrightarrow \mathbb{C}$ reads 
\begin{equation*}
\varphi (w)=\sqrt{n}w^{\frac{\alpha }{2}}.
\end{equation*}
where $\Sigma _{\alpha }(z)=\{w\in \mathbb{C}:\left\vert \arg z-\arg
w\right\vert <2\pi /\alpha \}$. The explicit relation between of integral
kernels reads 
\begin{equation}
\widehat{K}_{n}(z,w)=\varphi ^{\prime }(z)K(\varphi (z),\varphi (w))
\overline{\varphi ^{\prime }(w)}.
\end{equation}

Since $\varphi $ is a conformal transformation, we have  
\begin{equation*}
K(z,w)\overset{\varphi ^{-1}}{\longrightarrow }\widehat{K}_{n}^{\alpha }(z,w)
\end{equation*}
where $\varphi ^{-1}(w)=(w/n)^{2/\alpha }$ maps $\mathbb{C}$ into $\Sigma
_{\alpha }(z)$. Thus, we conclude that the family of potentials $V_{\alpha
}(z)=|z|^{\alpha }$ is conformally universal.

It is remarkable the fact that $\varphi ^{-1}$ is intimately connected to
unitary rescaling $g$. Introducing $u(w)=\sqrt{\alpha /2\pi }w$, we may
write 
\begin{equation*}
g(w)=(\varphi ^{-1}\circ u^{-1})(w),
\end{equation*}
where $u(w)$, acting as a rescaling of the complex plane, may stretch or
shrink the distances depending on $\alpha $. Hence, the unitary rescaling $g$
is associated with the universal behavior of the ensemble after a proper
rescaling of the complex plane.

Our results have also shown that for normal ensembles there is a distinction
between $\widetilde{K}_{\alpha}$ and the universal kernel $K(z,w)$.
Remarkably, the relation between $K$ and $\widetilde{K}_{\alpha }$ is
realized by a rescaling of the complex plane through the conformal map $u (
z ) =\sqrt{\frac{\alpha }{2\pi }}z.$ The explicit relation reads 
\begin{equation}
\widetilde{K}_{\alpha} ( z,w) = u ^{\prime }(z) K(u (z) , u ( w) )
\overline{u^{\prime }( w)}.
\end{equation}

The difference between $K$ and $\widetilde{K}^{\alpha }$ is not observed in the
Hermitian case. In the normal case the difference can be heuristically
explained by the difference in the dimension of the eigenvalues. For the
Hermitian case the influence of the Coulomb repulsion seems to be much more
relevant than for the normal ensemble. Moreover, the equilibrium disposition
of the eigenvalues varies with the parameter $\alpha$ due to the many
possible stable crystalline structures.

\textit{In summary,} we have shown that the concept of universality used for
Hermitian ensembles cannot be direct extrapolated to normal ensembles. We
have considered a model for random normal ensemble, where the potential is
given by a radially symmetric function depending only on one $parameter$. We
have shown that the rescaling of the average distance between the
eigenvalues to the unity, unlike for Hermitian ensembles, does not lead to a
universal eigenvalue statistics. To overcome this difficult, we have put
forward a new concept of universality for normal ensembles, which we called
conformal universality. We have shown that the rescaled integral kernel 
$K_{n}^{\alpha }(z,w)$ can be obtained by conformal transformations from a
universal kernel $K(z,w)$. Our result shows that the intricate relationship
between conformal geometry, revealed in the recent works 
\cite{Wiegman,Zabrodin}, might play an even more important role than previously
thought. It is wishful to extend the procedure adopted here to general
potentials. 

We are in debt with Prof. Dr. W. F. Wreszinski and D. B. Liarte for a critical 
and detailed reading of the manuscript. We would like to acknowledge the 
financial of the Brazilian agencies FAPESP (TP) and CNPq (AMV and 
DHUM).


\begin{thebibliography}{99}
\bibitem{comment} A major difference in the approach of Ref. \cite{Chau} and
ours is Eq. (\ref{Pm}). The authors consider a variant of it without
term $n$ multiplying Tr$[V(M)]$. In this case, the eigenvalues are
spread over the whole complex plane, and one cannot speak of eigenvalue
density in the limit $n \rightarrow \infty$. This subtle difference makes
the connection between the results extremely difficult.
\end{thebibliography}
\end{document}